\documentclass[12pt]{article}
\usepackage{amsmath,amssymb,amsthm,amsxtra,overpic,bbm,bm,epsfig,subfigure}
\usepackage{color}
\usepackage{comment}
\usepackage[all,cmtip]{xy}

\usepackage{slashed,stmaryrd}

\renewcommand{\thefootnote}{\fnsymbol{footnote}}

\begin{document}

\vspace{0.2cm}

\begin{center}
{\large\bf Another look at the impact of an eV-mass sterile neutrino on \\ the effective neutrino mass of neutrinoless double-beta decays}
\end{center}

\vspace{0.2cm}

\begin{center}
{\bf Jun-Hao Liu $^{a,~b}$} \footnote{E-mail: liujunhao@ihep.ac.cn}
\quad {\bf Shun Zhou $^{a,~b,~c}$} \footnote{E-mail: zhoush@ihep.ac.cn}
\\
{$^a$Institute of High Energy Physics, Chinese Academy of
Sciences, Beijing 100049, China \\
$^b$ School of Physical Sciences, University of Chinese Academy of Sciences, Beijing 100049, China\\
$^c$Center for High Energy Physics, Peking University, Beijing 100871, China}
\end{center}

\vspace{1.5cm}

\begin{abstract}
The possible existence of an eV-mass sterile neutrino, slightly mixing with ordinary active neutrinos, is not yet excluded by neutrino oscillation experiments. Assuming neutrinos to be Majorana particles, we explore the impact of such a sterile neutrino on the effective neutrino mass of neutrinoless double-beta decays $\langle m \rangle^\prime_{ee} \equiv m^{}_1 |V^{}_{e1}|^2 e^{{\rm i}\rho} + m^{}_2 |V^{}_{e2}|^2 + m^{}_3 |V^{}_{e3}|^2 e^{{\rm i}\sigma} + m^{}_4 |V^{}_{e4}|^2 e^{{\rm i}\omega}$, where $m^{}_i$ and $V^{}_{ei}$ (for $i = 1, 2, 3, 4$) denote respectively the absolute masses and the first-row elements of the 4$\times$4 neutrino flavor mixing matrix $V$, for which a full parametrization involves three Majorana-type CP-violating phases $\{\rho, \sigma, \omega\}$. A zero effective neutrino mass $|\langle m \rangle^\prime_{ee}| = 0$ is possible no matter whether three active neutrinos take the normal or inverted mass ordering, and its implications for the parameter space are examined in great detail. In particular, given the best-fit values of $m^{}_4 \approx 1.3~{\rm eV}$ and $|V^{}_{e4}|^2 \approx 0.019$ from the latest global analysis of neutrino oscillation data, a three-dimensional view of $|\langle m \rangle^\prime_{ee}|$ in the $(m^{}_1, \rho)$-plane is presented and further compared with that of the counterpart $|\langle m \rangle^{}_{ee}|$ in the absence of any sterile neutrino.
\end{abstract}

\begin{flushleft}
\hspace{0.88cm} PACS number(s): 14.60.Pq, 14.60.St, 25.30.Pt
\end{flushleft}

\def\thefootnote{\arabic{footnote}}
\setcounter{footnote}{0}

\newpage

Although Ettore Majorana noticed already 80 years ago~\cite{Majorana:1937vz} that neutrinos might be Majorana fermions, which are their own antiparticles, it is still unknown whether they are indeed so. The most promising way to demonstrate the Majorana nature of neutrinos is to observe the neutrinoless double-beta ($0\nu \beta \beta$) decays of an even-even atomic nucleus, i.e., $N(Z, A) \to N(Z+2, A) + 2 e^-$, where $Z$ and $A$ are respectively the atomic and mass numbers~\cite{Furry:1939qr}. Obviously, the $0\nu\beta\beta$ decays violate the lepton number by two units, and naturally happen when the electron (anti)neutrino flavor eigenstate is a linear superposition of three or more mass eigenstates of massive Majorana neutrinos~\cite{Rodejohann:2011mu, Bilenky:2012qi, Rodejohann:2012xd, Bilenky:2014uka, Pas:2015eia, DellOro:2016tmg}. In this case, the $0\nu\beta\beta$ decay rate is proportional to the square of the modulus of the effective neutrino mass
\begin{eqnarray}
\langle m \rangle^{}_{ee} \equiv \sum^3_{i=1} m^{}_i U^2_{ei} =  m^{}_1 |U^{}_{e1}|^2 e^{{\rm i}\rho} + m^{}_2 |U^{}_{e2}|^2 + m^{}_3 |U^{}_{e3}|^2 e^{{\rm i}\sigma} \; ,
\end{eqnarray}
in the scenario of three active neutrinos, where $m^{}_i$ and $U^{}_{ei}$ (for $i = 1, 2, 3$) are absolute neutrino masses and the first-row matrix elements of the mixing matrix $U$. If there exists an extra sterile neutrino of eV-mass, mixing slightly with active neutrinos, then the effective neutrino mass is
\begin{eqnarray}
\langle m \rangle^\prime_{ee} \equiv \sum^4_{i=1} m^{}_i V^2_{ei} =  m^{}_1 |V^{}_{e1}|^2 e^{{\rm i}\rho} + m^{}_2 |V^{}_{e2}|^2 + m^{}_3 |V^{}_{e3}|^2 e^{{\rm i}\sigma} + m^{}_4 |V^{}_{e4}|^2 e^{{\rm i}\omega} \; ,
\end{eqnarray}
where $m^{}_i$ and $V^{}_{ei}$ (for $i = 1, 2, 3, 4$) denote absolute neutrino masses and the first-row elements of the 4$\times$4 neutrino flavor mixing matrix $V$, respectively. Adopting the standard parametrization~\cite{Patrignani:2016xqp}, we can obtain $|V^{}_{e1}| = \cos \theta^{}_{14} \cos \theta^{}_{13} \cos \theta^{}_{12}$, $|V^{}_{e2}| = \cos \theta^{}_{14} \cos \theta^{}_{13} \sin \theta^{}_{12}$, $|V^{}_{e3}| = \cos \theta^{}_{14} \sin \theta^{}_{13}$ and $|V^{}_{e4}| = \sin \theta^{}_{14}$, where $\theta^{}_{ij}$ (for $ij = 12, 13, 14$) are neutrino mixing angles, and moreover three Majorana CP-violating phases $(\rho, \sigma, \omega)$ are involved. In the limit of $\theta^{}_{14} \to 0$, when the sterile neutrino is completely decoupled from ordinary active neutrinos, one can immediately reduce the four-flavor mixing matrix $V$ to the three-flavor mixing matrix $U$ and recover $\langle m\rangle^{}_{ee}$ from $\langle m\rangle^\prime_{ee}$.

The dependence of the $0\nu\beta\beta$ decays on the intrinsic properties of neutrinos can be perfectly described by the graphical representation of $|\langle m \rangle^{}_{ee}|$. For example, the most popular way is to show $|\langle m\rangle^{}_{ee}|$ for different values of the lightest neutrino mass in a two-dimensional plot~\cite{Vissani:1999tu}. In the case of normal neutrino mass ordering (NMO), namely, $m^{}_1 < m^{}_2 < m^{}_3$, $|\langle m \rangle^{}_{ee}|$ develops a vanishingly small value in the range of $2~{\rm meV} \lesssim m^{}_1 \lesssim 7~{\rm meV}$ due to an intricate cancellation caused by the Majorana CP-violating phases~\cite{Vissani:1999tu, Feruglio:2002af, Xing:2003jf}. In contrast, in the case of inverted neutrino mass ordering (IMO), i.e., $m^{}_3 < m^{}_1 < m^{}_2$, $|\langle m \rangle^{}_{ee}|$ can never be very small, and its lower bound $|\langle m \rangle^{}_{ee}| \gtrsim 15~{\rm meV}$ could promisingly be covered by the next-generation $0\nu\beta\beta$ decay experiments in the foreseeable future~\cite{KamLAND-Zen:2016pfg, Licciardi:2017oqg, Martin-Albo:2015rhw, Zhao:2016brs, Chen:2016qcd, Abgrall:2017syy}.

\begin{figure}[!t]
\begin{center}
\includegraphics[width=.7\textwidth]{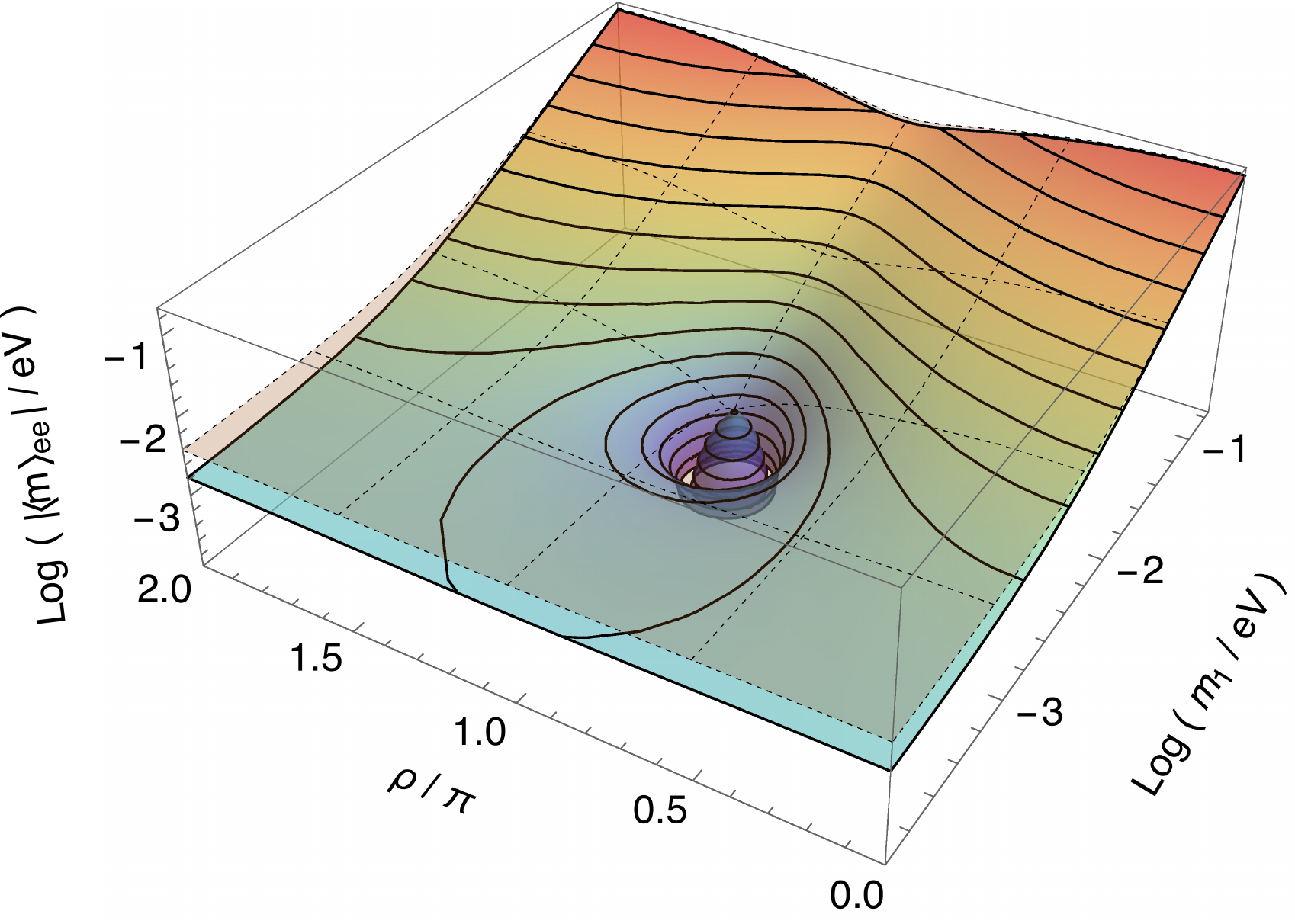} \\
\includegraphics[width=.7\textwidth]{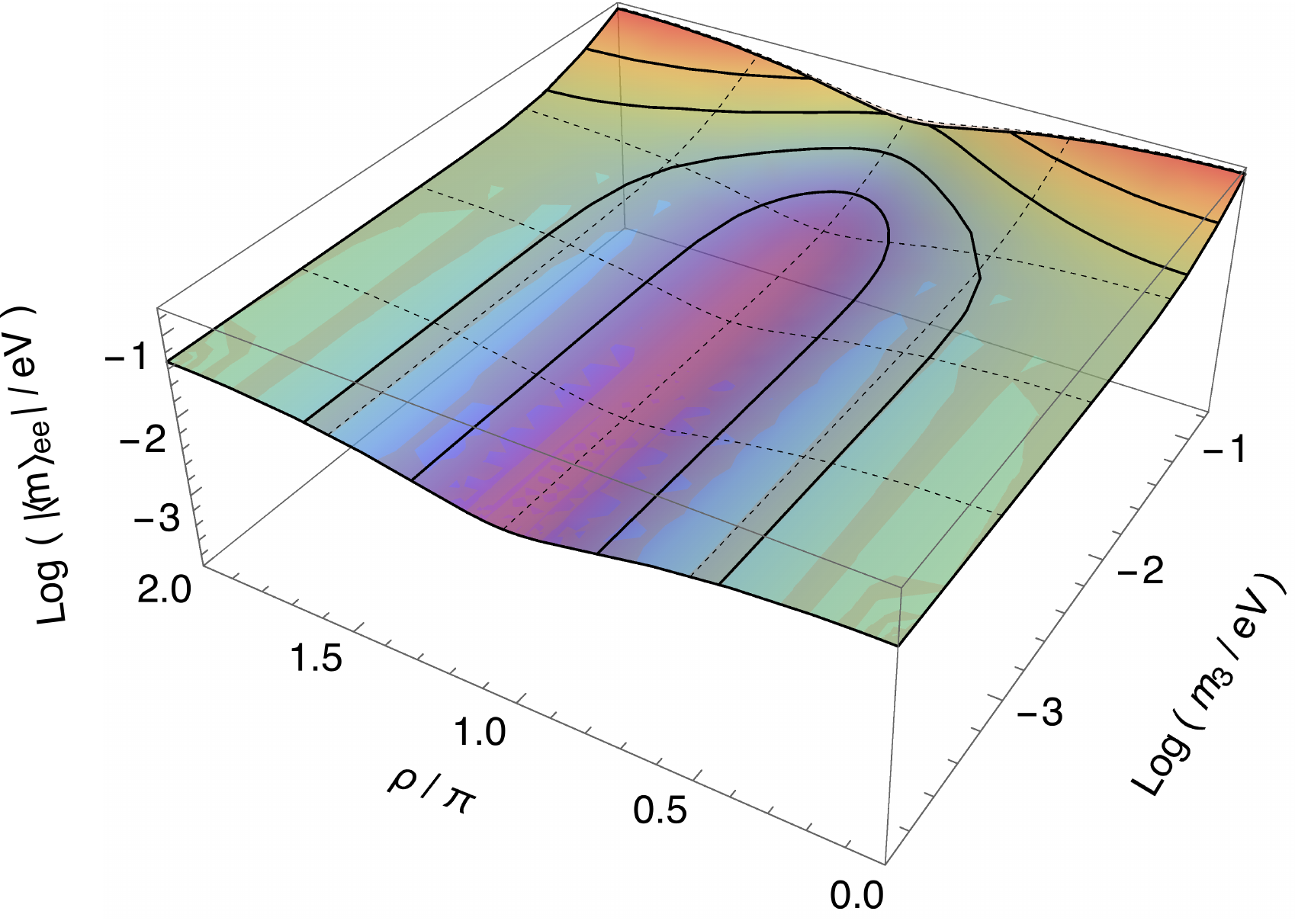}
\end{center}
\vspace{-0.4cm}
\caption{The upper and lower bounds of the effective neutrino mass $|\langle m \rangle^{}_{ee}|$ of the $0\nu\beta\beta$ decays in the framework of three ordinary active neutrinos, where the numerical results for NMO (IMO) are shown in the upper (lower) panel, as first presented in Ref.~\cite{Xing:2016ymd}. Here both plots are drawn with $\sin^2\theta^{}_{12} = 0.321$, $\sin^2 \theta^{}_{13} = 0.0216$, $\Delta m^2_{21} = 7.5\times 10^{-5}~{\rm eV}^2$ and $|\Delta m^2_{31}| = 2.5\times 10^{-3}~{\rm eV}^2$.
\label{fig:NOIO3}}
\end{figure}

More interestingly, it has been found that a three-dimensional plot of $|\langle m \rangle^{}_{ee}|$ with respect to $\rho$ and $m^{}_1$ in the NMO case reveals a bullet-like structure in the region where the cancellation occurs and can even lead to $|\langle m \rangle^{}_{ee}| = 0$~\cite{Xing:2015zha, Xing:2016ymd}. In particular, the authors of Ref.~\cite{Xing:2016ymd} have discovered that the surfaces of upper and lower limits of $|\langle m \rangle^{}_{ee}|$, through varying the other Majorana CP-violating phase $\sigma$, have a contact point at $(\rho = \pi, m^{}_1 = 4~{\rm meV}, |\langle m \rangle^{}_{ee}| = 1~{\rm meV})$. It is suggested that $|\langle m\rangle^*_{ee}| = 1~{\rm meV}$ may be regarded as a target value for the sensitivity of future experiments in case that NMO is verified by the forthcoming neutrino oscillation experiments~\cite{Xing:2016ymd}. This suggestion is justified by the fact that the probability for $|\langle m \rangle^{}_{ee}|$ to be lying below $|\langle m\rangle^*_{ee}| = 1~{\rm meV}$ is tiny, given the latest neutrino oscillation data and the cosmological bound on neutrino masses~\cite{Benato:2015via,Ge:2016tfx, Caldwell:2017mqu, Agostini:2017jim, Ge:2017erv}. For illustration, we have reproduced the three-dimensional graph of $|\langle m \rangle^{}_{ee}|$ in Fig.~\ref{fig:NOIO3}, where both results for NMO and IMO are given.

In this short note, we follow the idea of Ref.~\cite{Xing:2016ymd} for the three-flavor mixing of active neutrinos and generalize it to the so-called $3+1$ active-sterile neutrino mixing scenario, in which an extra sterile neutrino of eV-mass is introduced and it slightly mixes with three ordinary neutrinos. The motivation for such a generalization is two-fold. First, an eV-mass sterile neutrino has recently received a lot of attention because of the tentative anomalies arising from the short-baseline reactor neutrino experiments~\cite{Gariazzo:2015rra}. Although the combined analysis of Daya Bay, MINOS and Bugey-3 data~\cite{Adamson:2016jku} has excluded most of the parameter space of the mass-squared difference $\Delta m^2_{41} \equiv m^2_4 - m^2_1$ and the mixing angle $\sin^2 \theta^{}_{14}$ for the sterile neutrino, the latest global analysis of neutrino oscillation data indicates that a small region around the best-fit values of $\Delta m^2_{41} = 1.7~{\rm eV}^2$ and $\sin^2 \theta^{}_{14} = 0.019$ is still allowed~\cite{Gariazzo:2017fdh}. The existence of an eV-mass sterile neutrino has very important implications for the effective neutrino mass of $0\nu\beta\beta$ decays, which have been extensively discussed in the literature~\cite{Goswami:2005ng, Goswami:2007kv, Barry:2011wb, Li:2011ss, Girardi:2013zra, Giunti:2015kza}. Therefore, it is interesting to see how the fine structure of $|\langle m\rangle^{}_{ee}|$ observed in the three-flavor neutrino mixing scenario is modified in the presence of a sterile neutrino. Second, the analysis of $\langle m\rangle^\prime_{ee}$ can be viewed as a representative example for new physics beyond the standard mechanism for the $0\nu\beta\beta$ decays. If the contributions from new physics are similar to those from light neutrinos, while the relevant nuclear matrix element is unchanged, they can be described by the additional term in the effective neutrino mass $\langle m\rangle^\prime_{ee}$~\cite{Xing:2015zha}. In this sense, the conclusions for the case of eV-mass sterile neutrinos will be equally applicable to some other new physics scenarios.

\vspace{0.5cm}

First of all, let us summarize the main features of $|\langle m\rangle^{}_{ee}|$ in the three-flavor neutrino mixing case, and provide some new insights into the bullet-like structure appearing in the NMO case. Following Ref.~\cite{Xing:2016ymd}, we define the sum of the first two terms in $\langle m\rangle^{}_{ee}$ as $m^{}_{12} \cos^2 \theta^{}_{13}$, where
\begin{eqnarray}
m^{}_{12} \equiv m^{}_1 \cos^2\theta^{}_{12} e^{{\rm i}\rho} + m^{}_2 \sin^2 \theta^{}_{12} \; ,
\end{eqnarray}
and the standard parametriztion $|U^{}_{e1}| = \cos \theta^{}_{13} \cos \theta^{}_{12}$, $|U^{}_{e2}| = \cos \theta^{}_{13} \sin \theta^{}_{12}$ and $|U^{}_{e3}| = \sin \theta^{}_{13}$ have been used~\cite{Patrignani:2016xqp}. We further express $m^{}_{12} = \overline{m}^{}_{12} e^{{\rm i}\alpha^{}_{12}}$ in term of its modulus $\overline{m}^{}_{12} \equiv |m^{}_{12}|$ and its argument $\alpha^{}_{12}$, which can be extracted from $\tan \alpha^{}_{12} = m^{}_1\sin\rho/(m^{}_2 \tan^2\theta^{}_{12} + m^{}_1\cos\rho)$. Then, it is straightforward to observe that the upper and lower bounds of $|\langle m\rangle^{}_{ee}|$ are given by the constructive and destructive interference between $\overline{m}^{}_{12} \cos^2\theta^{}_{13}e^{{\rm i}\alpha^{}_{12}}$ and $m^{}_3 \sin^2 \theta^{}_{13} e^{{\rm i}\sigma}$, corresponding to their phase difference being $0$ and $\pm\pi$, respectively. To be explicit, we have the upper and lower bounds of $|\langle m\rangle^{}_{ee}|$ as functions of $m^{}_1$ and $\rho$~\cite{Xing:2016ymd}
\begin{eqnarray}
|\langle m\rangle^{}_{ee}|^{}_{\rm U, L} = |\overline{m}^{}_{12} \cos^2\theta^{}_{13} \pm m^{}_3 \sin^2\theta^{}_{13}| \; ,
\end{eqnarray}
where the meaning of the subscripts ``U" and ``L" is self-evident and
\begin{eqnarray}
\overline{m}^{}_{12} \equiv \sqrt{m^2_1 \cos^4\theta^{}_{12} + 2 m^{}_1 m^{}_2 \sin^2\theta^{}_{12} \cos^2\theta^{}_{12} \cos\rho + m^2_2 \sin^4\theta^{}_{12}} \; .
\end{eqnarray}
Given two mixing angles $\sin^2\theta^{}_{12} = 0.321$ and $\sin^2 \theta^{}_{13} = 0.0216$, and two neutrino mass-squared differences $\Delta m^2_{21} \equiv m^2_2 - m^2_1 = 7.5\times 10^{-5}~{\rm eV}^2$ and $\Delta m^2_{31} \equiv m^2_3 - m^2_1 = 2.5\times 10^{-3}~{\rm eV}^2$ from neutrino oscillation data, one can figure out $m^{}_2 = \sqrt{m^2_1 + \Delta m^2_{21}}$ and $m^{}_3 = \sqrt{m^2_1 + \Delta m^2_{31}}$ and then draw $|\langle m\rangle^{}_{ee}|^{}_{\rm U}$ and $|\langle m\rangle^{}_{ee}|^{}_{\rm L}$ as functions of two unknown parameters $m^{}_1$ and $\rho$. In Fig.~\ref{fig:NOIO3}, the numerical results of $|\langle m\rangle^{}_{ee}|^{}_{\rm U}$ and $|\langle m\rangle^{}_{ee}|^{}_{\rm L}$ are shown in the upper panel. The upper bound $|\langle m\rangle^{}_{ee}|^{}_{\rm U}$ is plotted in light pink, while the lower bound $|\langle m\rangle^{}_{ee}|^{}_{\rm L}$ in gradient color with black solid contours of a few typical values of $|\langle m\rangle^{}_{ee}|$. Some comments are in order:
\begin{itemize}
\item As found in Ref.~\cite{Xing:2016ymd}, the local minimum of $|\langle m\rangle^{}_{ee}|^{}_{\rm U}$ and the local maximum of $|\langle m\rangle^{}_{ee}|^{}_{\rm L}$ are achieved at the same point $m^{}_1 = m^{}_2 \tan^2\theta^{}_{12} = \sqrt{\Delta m^2_{21}} \sin^2\theta^{}_{12}/\sqrt{\cos 2\theta^{}_{12}} \approx 4.65~{\rm meV}$ and $\rho = \pi$, and they have a contact at $|\langle m\rangle^*_{ee}| = m^{}_3\sin^2\theta^{}_{13} \approx 1~{\rm meV}$. These features can be well understood by noticing that the functional properties of both the upper and lower bounds are determined by the same function $\overline{m}^{}_{12}$, whose definition is given in Eq.~(5). It is easy to see that it cannot be negative and the minimum $\overline{m}^{}_{12} = 0$ is reached at $\rho = \pi$ and $m^{}_1 = m^{}_2 \tan^2\theta^{}_{12}$. Since $|\langle m\rangle^{}_{ee}|^{}_{\rm U} = \overline{m}^{}_{12} \cos^2 \theta^{}_{13} + m^{}_3 \sin^2\theta^{}_{13}$ is always positive, its local minimum $+m^{}_3 \sin^2\theta^{}_{13}$ is achieved at $\overline{m}^{}_{12} = 0$. On the other hand, looking at $\overline{m}^{}_{12} \cos^2 \theta^{}_{13} - m^{}_3 \sin^2\theta^{}_{13}$, we can observe that the local minimum $- m^{}_3 \sin^2\theta^{}_{13}$ of itself is also achieved at $\overline{m}^{}_{12} = 0$. However, after taking the absolute value, it becomes the local maximum of $|\langle m\rangle^{}_{ee}|^{}_{\rm L}$. Furthermore, the condition for $|\langle m\rangle^{}_{ee}|^{}_{\rm L} = 0$ is given by $\overline{m}^{}_{12} = m^{}_3\tan^2 \theta^{}_{13}$, from which one can determine the closed curve at the bottom of the bullet in the $(\rho, m^{}_1)$-plane.

\item In the lower panel of Fig.~\ref{fig:NOIO3}, the numerical results of $|\langle m\rangle^{}_{ee}|^{}_{\rm U}$ and $|\langle m\rangle^{}_{ee}|^{}_{\rm L}$ are also given in the IMO case. In this case, $m^{}_3$ is the lightest neutrino mass, implying $m^{}_1 = \sqrt{m^2_3 - \Delta m^2_{31}}$ and $m^{}_2 = \sqrt{m^2_3 - \Delta m^2_{31} + \Delta m^2_{21}}$, where $\Delta m^2_{31} = - 2.5\times 10^{-3}~{\rm eV}^2$ and $\Delta m^2_{21} = 7.5\times 10^{-5}~{\rm eV}^2$. Given the definitions of $|\langle m\rangle^{}_{ee}|^{}_{\rm U}$ and $|\langle m\rangle^{}_{ee}|^{}_{\rm L}$ in Eq.~(4), we can immediately find that the difference between them is extremely small, namely, $m^{}_3 \sin^2\theta^{}_{13} \lesssim 2~{\rm meV}$ even for $m^{}_3 = 0.1~{\rm eV}$. This is the reason why two surfaces for $|\langle m\rangle^{}_{ee}|^{}_{\rm U}$ and $|\langle m\rangle^{}_{ee}|^{}_{\rm L}$ almost overlap with each other. Furthermore, neglecting the contribution from $m^{}_3 \sin^2\theta^{}_{13}$, we observe that the minima of $|\langle m\rangle^{}_{ee}|^{}_{\rm U}$ and $|\langle m\rangle^{}_{ee}|^{}_{\rm L}$ are given by the same condition ${\rm d}\overline{m}^2_{12}/{\rm d}\rho \propto \sin\rho = 0$ and can be achieved at $\rho = \pi$. Unlike the NMO case, the condition $m^{}_1 \approx 4.65~{\rm meV}$ is no longer allowed in the IMO case. For instance, we have $m^{}_2 \approx m^{}_1 \approx 50~{\rm meV}$ even for $m^{}_3 = 0$. Hence there will be no significant cancellation in $|\langle m\rangle^{}_{ee}|$ for IMO, as is well known~\cite{Bilenky:2014uka}.
\end{itemize}
While $|\langle m\rangle^{}_{ee}| = 0$ is impossible for three active neutrinos in the IMO case, as we shall show below, there is an interesting fine structure of $|\langle m\rangle^\prime_{ee}|$ in the presence of a sterile neutrino even in the IMO case. In the NMO case, the fine structure will also be dramatically modified.

\vspace{0.5cm}

Then, we proceed with the effective neutrino mass $\langle m\rangle^\prime_{ee}$ for $0\nu\beta\beta$ decays in the presence of an eV-mass sterile neutrino, which has been defined in Eq.~(2). In our discussions, we use only the best-fit values of $\Delta m^2_{41} = 1.7~{\rm eV}^2$ and $\sin^2\theta^{}_{14} = 0.019$, which happen to be lying in the region where the contributions from active neutrinos and the sterile neutrino are comparable. In addition, the sterile neutrino mass $m^{}_4$ is taken to be heaviest such that $m^{}_4 \approx \sqrt{\Delta m^2_{41}} \approx 1.3~{\rm eV}$ and the mass ordering of three active neutrinos is the same as in the standard case. Inspired by the previous analysis in the three-flavor neutrino mixing scenario, we introduce~\footnote{Note that $m^\prime_{123}$ can actually be identified as the effective neutrino mass $\langle m\rangle^{}_{ee}$ in the three-flavor neutrino mixing scenario. The global-fit results of the mixing parameters are in general different for three-flavor and four-flavor mixing scenarios, but here we take the same input values just for illustration. To avoid any confusion, we hereafter use $m^\prime_{123}$ instead of $\langle m\rangle^{}_{ee}$.}
\begin{eqnarray}
m^\prime_{123} \equiv m^{}_{12} \cos^2\theta^{}_{13} + m^{}_3 \sin^2\theta^{}_{13}e^{{\rm i}\sigma} \; ,
\end{eqnarray}
where $m^{}_{12}$ is given in Eq.~(3), and derive the upper and lower bounds of $|\langle m\rangle^\prime_{ee}|$ as follows
\begin{eqnarray}
|\langle m\rangle^\prime_{ee}|^{}_{\rm U, L} = |\overline{m}^\prime_{123} \cos^2\theta^{}_{14} \pm m^{}_4 \sin^2\theta^{}_{14}| \; ,
\end{eqnarray}
with
\begin{eqnarray}
\overline{m}^\prime_{123} = \sqrt{\overline{m}^2_{12} \cos^4\theta^{}_{13} + 2\overline{m}^{}_{12} m^{}_3 \sin^2\theta^{}_{13}\cos^2\theta^{}_{13} \cos(\alpha^{}_{12} - \sigma) + m^2_3 \sin^4 \theta^{}_{13}} \; .
\end{eqnarray}

\begin{figure}[t!]
\begin{center}
\includegraphics[width=.7\textwidth]{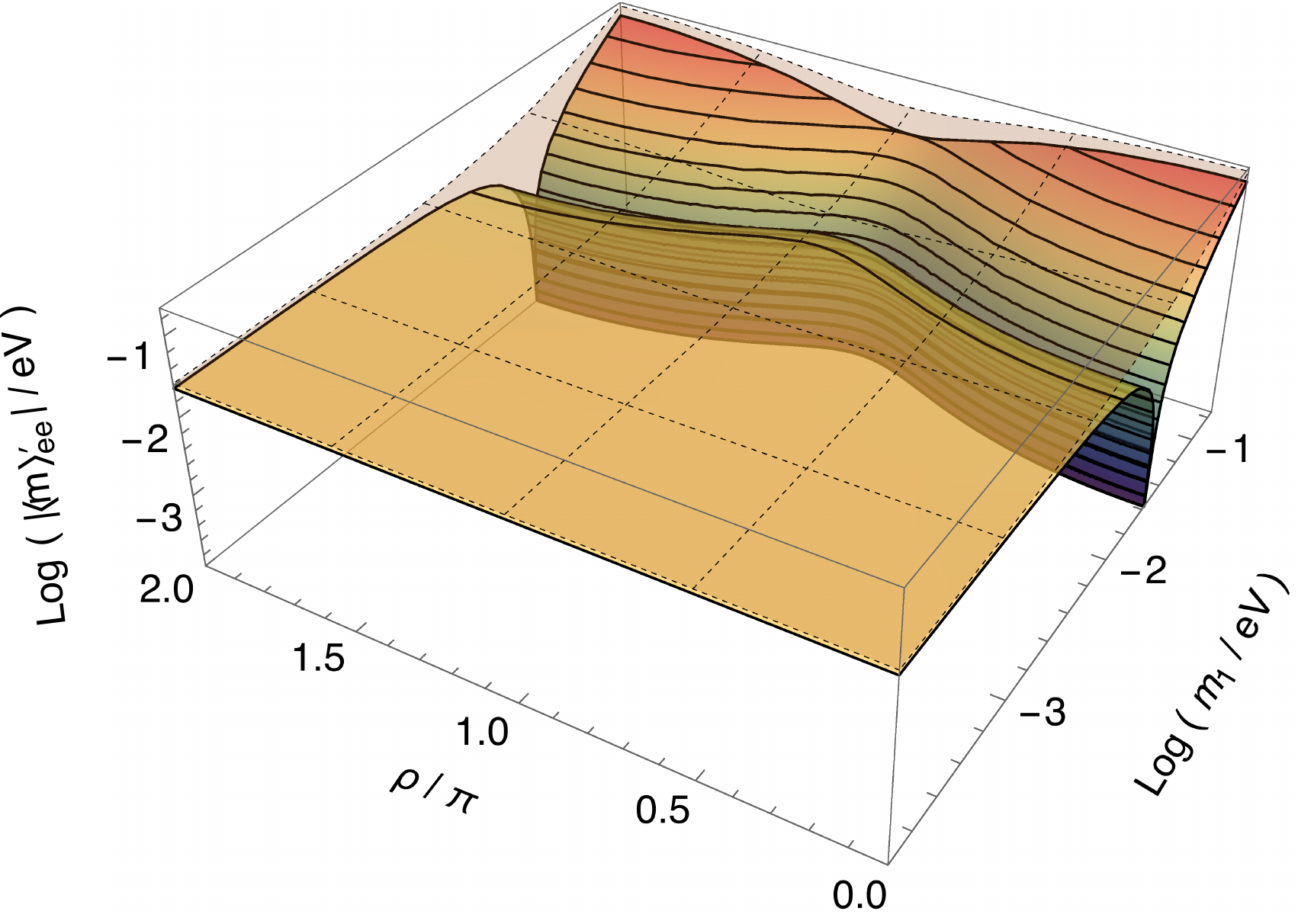}\\
\includegraphics[width=.7\textwidth]{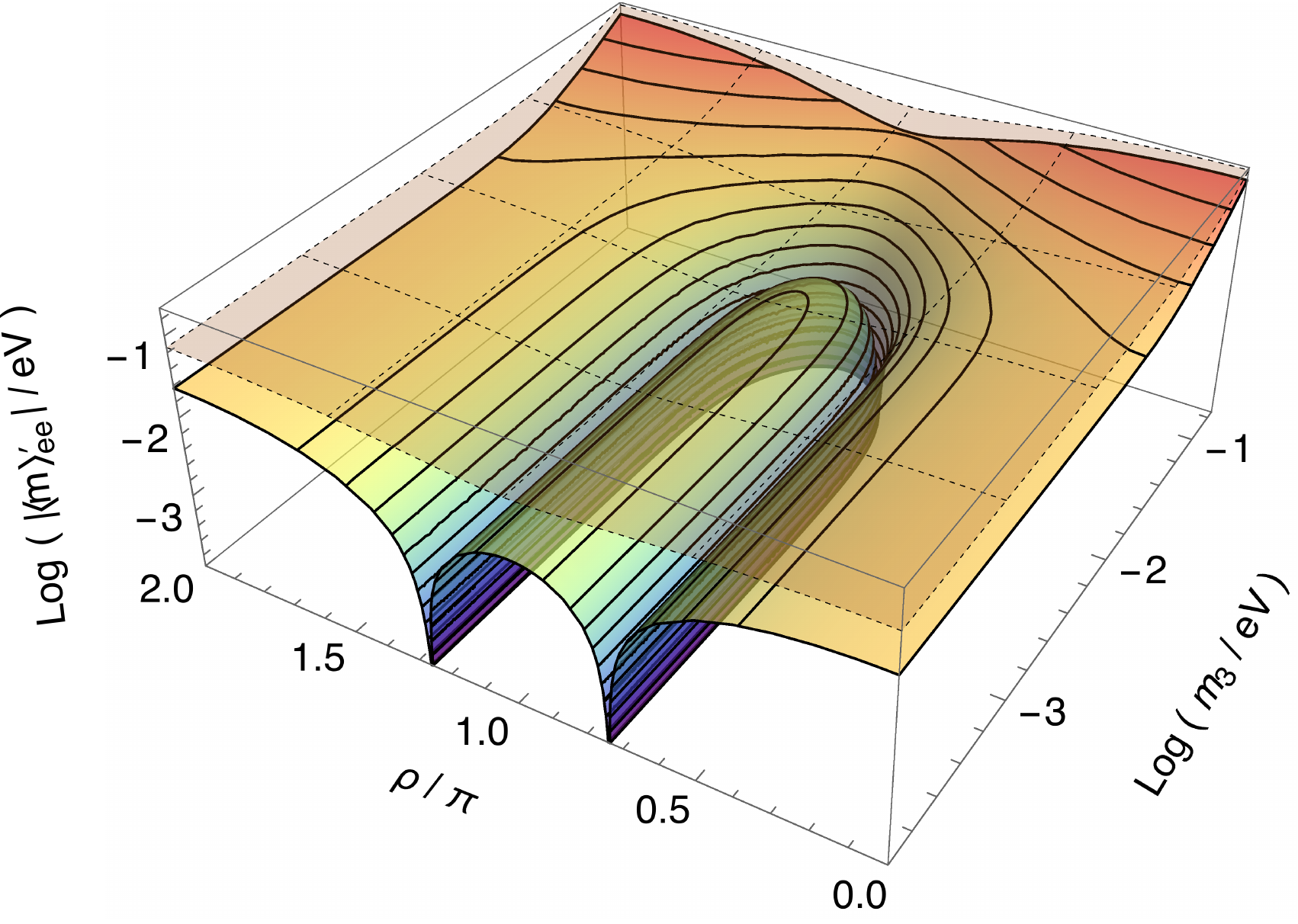}
\end{center}
\vspace{-0.4cm}
\caption{The upper and lower bounds of the effective neutrino mass $|\langle m \rangle^\prime_{ee}|$ of the $0\nu\beta\beta$ decays in the framework of $3+1$ active-sterile neutrino mixing, where the numerical results for the NMO (IMO) of three ordinary neutrinos are shown in the upper (lower) panel. We adopt the input values $\Delta m^2_{41} = 1.7~{\rm eV}^2$, $\sin^2\theta^{}_{14} = 0.019$ and $\sigma = \pi$, while the others are the same as in Fig.~\ref{fig:NOIO3}.
\label{fig:NOIO4}}
\end{figure}

As before, it is straightforward to observe that the upper and lower bounds correspond to the phase difference between $m^\prime_{123}\cos^2 \theta^{}_{14}$ and $m^{}_4 \sin^2 \theta^{}_{14} e^{{\rm i}\omega}$ being $0$ and $\pm\pi$, respectively. In Fig.~\ref{fig:NOIO4}, we show the numerical results of $|\langle m\rangle^\prime_{ee}|^{}_{\rm U}$ and $|\langle m\rangle^\prime_{ee}|^{}_{\rm L}$ as functions of $\rho$ and $m^{}_1$ in the upper panel for NMO, while as functions of $\rho$ and $m^{}_3$ in the lower panel for IMO. It is evident that $|\langle m\rangle^\prime_{ee}|$ can be vanishing for both mass orderings of three active neutrinos. Now we discuss the main features of $|\langle m\rangle^\prime_{ee}|^{}_{\rm U}$ and $|\langle m\rangle^\prime_{ee}|^{}_{\rm L}$ by following an approximate and analytical approach. Since the upper bound $|\langle m\rangle^\prime_{ee}|^{}_{\rm U}$ is always positive, we focus on the lower bound $|\langle m\rangle^\prime_{ee}|^{}_{\rm L}$.

Let us explore the conditions under which a vanishing effective neutrino mass $|\langle m\rangle^\prime_{ee}|^{}_{\rm L} = 0$ can be achieved. Since there are many free parameters involved in the effective neutrino mass, we fix the mixing parameters $\sin^2 \theta^{}_{12} = 0.321$, $\sin^2 \theta^{}_{13} = 0.0216$, $\Delta m^2_{21} = 7.5\times 10^{-5}~{\rm eV}^2$, $|\Delta m^2_{31}| = 2.5\times 10^{-3}~{\rm eV}^2$ for active neutrinos, and $\Delta m^2_{41} = 1.7~{\rm eV}^2$ and $\sin^2 \theta^{}_{14} = 0.019$ for the sterile neutrino. The free parameters in question include the lightest neutrino mass $m^{}_1$ for NMO ($m^{}_3$ for IMO), and three Majorana CP-violating phases $(\rho, \sigma, \omega)$. To obtain $|\langle m\rangle^\prime_{ee}|^{}_{\rm L} = 0$, one can immediately infer from Eq.~(7) that the equality below should be satisfied
\begin{eqnarray}
\overline{m}^\prime_{123} = m^{}_4 \tan^2\theta^{}_{14}\; ,
\end{eqnarray}
where $m^{}_4 \tan^2 \theta^{}_{14} \approx \sqrt{\Delta m^2_{41}} \tan^2 \theta^{}_{14} \approx 25.3~{\rm meV}$ holds as a good approximation due to $m^{}_4 \gg m^{}_1$. As we have seen from the previous discussions, the fine structure of $|\langle m\rangle^{}_{ee}| = \overline{m}^\prime_{123}$ appears in the region of cancellation and the local maximum of $|\langle m\rangle^{}_{ee}|^{}_{\rm L}$ is $|\langle m\rangle^*_{ee}| = 1~{\rm meV}$, which is much smaller than $m^{}_4\tan^2 \theta^{}_{14} \approx 25.3~{\rm meV}$. Therefore, the equality in Eq.~(9) can only be satisfied in the parameter space where a relatively large value of $\overline{m}^\prime_{123}$ or $|\langle m\rangle^{}_{ee}| \approx 25.3~{\rm meV}$ is obtained. Such a large value can be realized only for a sizable $m^{}_1$ in the NMO case, but this is not necessary for $m^{}_3$ in the IMO case. In the following, we separately analyze the solutions for NMO and IMO.
\begin{itemize}
\item {\it The NMO Case} --- Since we are far from the parameter space, where the fine structure of $|\langle m \rangle^{}_{ee}|$ is located, the difference between $|\langle m \rangle^{}_{ee}|^{}_{\rm U}$ and $|\langle m \rangle^{}_{ee}|^{}_{\rm L}$ is as small as $m^{}_3 \sin^2\theta^{}_{13} = 2.19~{\rm meV}$ even for $m^{}_3 = 0.1~{\rm eV}$. Consequently, the dependence of the desired conditions on the Majorana phase $\sigma$ is very weak, and one can further arrive at
    \begin{eqnarray}
    m^2_1 \cos^4 \theta^{}_{12} + m^2_2 \sin^4 \theta^{}_{12} + \frac{1}{2} m^2_1 m^2_2 \sin^2 2\theta^{}_{12} \cos \rho = m^2_4 \tan^4\theta^{}_{14} \sec^4 \theta^{}_{13} \; ,
    \end{eqnarray}
    which stands for a curve in the $(\rho, m^{}_1)$-plane. Along this curve, $|\langle m\rangle^\prime_{ee}| = 0$ is achieved. As indicated in the bottom plane of the top panel of Fig.~\ref{fig:NOIO4}, this takes place only for a small interval of $m^{}_1$, but the whole range of $\rho$. Notice that the left-hand side of Eq.~(10) reaches its maximum for $\rho = 0$ or $2\pi$ and its minimum for $\rho = \pi$, given a value of $m^{}_1$. To find out this interval, we ignore the tiny difference between $m^{}_1$ and $m^{}_2$, and solve Eq.~(10) for the minimum $m^{\rm min}_1$ by setting $\rho = 0$ or $2\pi$, namely,
    \begin{eqnarray}
    m^{\rm min}_1 \approx m^{}_4 \tan^2 \theta^{}_{14} \sec^2\theta^{}_{13} \approx 25.8~{\rm meV} \; .
    \end{eqnarray}
    In a similar way, we can derive the maximum $m^{\rm max}_1$ by taking $\rho = \pi$, i.e.,
    \begin{eqnarray}
    m^{\rm max}_1 \approx m^{}_4\tan^2\theta^{}_{14} \sec^2 \theta^{}_{13}/\cos2\theta^{}_{12} \approx 72.1~{\rm meV} \; .
    \end{eqnarray}
    Hence the curve for $|\langle m\rangle^\prime_{ee}| = 0$ appears within $25.8~{\rm meV} \lesssim m^{}_1 \lesssim 72.1~{\rm meV}$ and $0 \leq \rho \leq 2\pi$.

    Moreover, it is also interesting to point out that $|\langle m\rangle^\prime_{ee}|^{}_{\rm U}$ and $|\langle m\rangle^\prime_{ee}|^{}_{\rm L}$ have a contact point, which is very much similar to the case of three-flavor mixing. Looking at Eq.~(7), one can recognize that if $\overline{m}^\prime_{123} = 0$ is allowed, then $|\langle m\rangle^\prime_{ee}|^{}_{\rm U} = |\langle m\rangle^\prime_{ee}|^{}_{\rm L} = m^{}_4 \sin^2\theta^{}_{14} \approx 24.8~{\rm meV}$. With the help of Eq.~(8), requiring $\overline{m}^\prime_{123} = 0$, we get two familiar conditions $\alpha^{}_{12} = \sigma \pm \pi$ and $\overline{m}^{}_{12} = m^{}_3 \tan^2 \theta^{}_{13}$, which are the same for $|\langle m\rangle^{}_{ee}| = 0$ for the standard case of three-flavor mixing and have been extensively discussed in Refs.~\cite{Xing:2015zha,Xing:2016ymd,Benato:2015via}. Therefore, the location of the contact point in the $(\rho, m^{}_1)$-plane depends on the exact value of $\sigma$. When $\sigma$ is varying from $0$ to $2\pi$, the corresponding $(\rho, m^{}_1)$ coordinates of the contact point are moving along the closed curve determined by $|\langle m\rangle^{}_{ee}| =0$.

\item {\it The IMO Case} --- Note that the upper and lower bounds of $|\langle m\rangle^{}_{ee}|$ are quite similar for the whole range of $m^{}_3 \lesssim 0.1~{\rm eV}$.  As a consequence, the equalities in Eqs.~ (9) and (10) are still applicable in this case. However, two different scenarios of small and large values of $m^{}_3$ should be distinguished. In the former case of $m^{}_3 \ll \sqrt{|\Delta m^2_{31}|} = 50~{\rm meV}$, we have $m^{}_2 \approx m^{}_1 = \sqrt{m^2_3 + |\Delta m^2_{31}|} \approx 50~{\rm meV}$, which is larger than $m^{}_4 \tan^2\theta^{}_{14} \sec^2\theta^{}_{13} \approx 25.8~{\rm meV}$, implying $\cos\rho < 0$. More explicitly, we obtain
    \begin{eqnarray}
    \cos\rho = \frac{m^2_4\tan^4\theta^{}_{14} \sec^4 \theta^{}_{13} - (m^2_1\cos^4 \theta^{}_{12} + m^2_2 \sin^4 \theta^{}_{12})}{2m^{}_1 m^{}_2 \sin^2 \theta^{}_{12} \cos^2 \theta^{}_{12}} \approx -0.683 \; ,
    \end{eqnarray}
    or equivalently $\rho = 0.74~\pi$ or $1.26~\pi$ in the limit of $m^{}_3 = 0$. In the latter case of $m^{}_3 \gg 50~{\rm meV}$, one can get $m^{}_2 \approx m^{}_1 = \sqrt{m^2_3 + |\Delta m^2_{31}|} \approx m^{}_3$. In this case, for $\rho = \pi$, Eq.~(10) leads to
    \begin{eqnarray}
    m^{\rm max}_3 \approx m^{}_4\tan^2\theta^{}_{14} \sec^2 \theta^{}_{13}/\cos2\theta^{}_{12} \approx 72.1~{\rm meV} \; ,
    \end{eqnarray}
    which turns out to be comparable to $\sqrt{|\Delta m^2_{31}|} = 50~{\rm meV}$. Therefore, there exists an upper bound on $m^{}_3 < m^{\rm max}_3$, above which it is impossible to get $|\langle m\rangle^\prime_{ee}| = 0$. As shown in the bottom plane of the lower panel of Fig.~\ref{fig:NOIO4}, the curve in the $(\rho, m^{}_3)$-plane for $|\langle m\rangle^\prime_{ee}| = 0$ is constrained to be within $0.74~\pi \lesssim \rho \lesssim 1.26~\pi$ and $0 < m^{}_3 < 72.1~{\rm meV}$.
\end{itemize}
Although the above analysis is approximate to some extent and has been performed for given values of $\Delta m^2_{41}$ and $\sin^2\theta^{}_{14}$, it serves as a typical example to understand the main features of the effective neutrino mass $\langle m\rangle^\prime_{ee}$ for $0\nu\beta\beta$ decays in the presence of a sterile neutrino.

\vspace{0.5cm}

Finally, we give some remarks on the $0\nu\beta\beta$ decays and summarize our main conclusions. Let us tentatively assume that there is no new physics beyond the standard model other than three massive neutrinos, which are required to be responsible for neutrino oscillations. Then, it would be perfect that the future neutrino oscillation experiments tell us that neutrino mass ordering is inverted, while the next-generation experiments observe $0\nu\beta\beta$ decays with a half-life implied by an effective neutrino mass in the right region. In such an optimistic case, we will definitely know that neutrinos are Majorana particles~\cite{Schechter:1981bd, Duerr:2011zd, Liu:2016oph}. On the other hand, if the $0\nu\beta\beta$ decays are not observed, neutrinos should be Dirac particles. However, if we allow new physics to contribute to $0\nu\beta\beta$ decays, e.g., an eV-mass sterile neutrino mixing with active neutrinos, neutrinos can be Majorana particles even in the case where IMO is true and $0\nu\beta\beta$ decays are absent.

In the present short note, we have explored the impact of a sterile neutrino on the effective neutrino mass $\langle m\rangle^\prime_{ee}$. Taking the current best-fit values of $\Delta m^2_{41} = 1.7~{\rm eV}^2$ and $\sin^2 \theta^{}_{14} = 0.019$, we find that the contributions from the sterile neutrino are comparable to those from active neutrinos in both NMO and IMO cases, implying an intriguing interplay between them. The conditions for $|\langle m\rangle^\prime_{ee}| = 0$ and the fine structures in this region, which are quite different from the standard case, are studied analytically and illustrated in three-dimensional plots as well. Our results will be helpful in understanding the possible effects on $0\nu\beta\beta$ decays from sterile neutrinos or some other new physics. As the very short baseline neutrino oscillation experiments will hopefully make a final verdict on the existence of sterile neutrinos soon, the outcomes of future $0\nu\beta\beta$ decay experiments are crucially important for probing the intrinsic properties of massive neutrinos, such as the Majorana nature and the Majorana CP-violating phases.

\section*{Acknowledgements}

This work was supported in part by the National Recruitment Program for Young Professionals and by the CAS Center for Excellence in Particle Physics (CCEPP).


\begin{thebibliography}{99}
\bibitem{Majorana:1937vz}
  E.~Majorana,
  ``Teoria simmetrica dell¡¯elettrone e del positrone,''
  Nuovo Cim.\  {\bf 14}, 171 (1937).

\bibitem{Furry:1939qr}
  W.~H.~Furry,
  ``On transition probabilities in double beta-disintegration,''
  Phys.\ Rev.\  {\bf 56}, 1184 (1939).

\bibitem{Rodejohann:2011mu}
  W.~Rodejohann,
  ``Neutrino-less Double Beta Decay and Particle Physics,''
  Int.\ J.\ Mod.\ Phys.\ E {\bf 20}, 1833 (2011)
  [arXiv:1106.1334].

\bibitem{Bilenky:2012qi}
  S.~M.~Bilenky and C.~Giunti,
  ``Neutrinoless double-beta decay: A brief review,''
  Mod.\ Phys.\ Lett.\ A {\bf 27}, 1230015 (2012)
  [arXiv:1203.5250].

\bibitem{Rodejohann:2012xd}
  W.~Rodejohann,
  ``Neutrinoless double beta decay and neutrino physics,''
  J.\ Phys.\ G {\bf 39}, 124008 (2012)
  [arXiv:1206.2560].

\bibitem{Bilenky:2014uka}
  S.~M.~Bilenky and C.~Giunti,
  ``Neutrinoless Double-Beta Decay: a Probe of Physics Beyond the Standard Model,''
  Int.\ J.\ Mod.\ Phys.\ A {\bf 30}, no. 04n05, 1530001 (2015)
  [arXiv:1411.4791].

\bibitem{Pas:2015eia}
  H.~P\"{a}s and W.~Rodejohann,
  ``Neutrinoless Double Beta Decay,''
  New J.\ Phys.\  {\bf 17}, no. 11, 115010 (2015)
  [arXiv:1507.00170].

\bibitem{DellOro:2016tmg}
  S.~Dell'Oro, S.~Marcocci, M.~Viel and F.~Vissani,
  ``Neutrinoless double beta decay: 2015 review,''
  Adv.\ High Energy Phys.\  {\bf 2016}, 2162659 (2016)
  [arXiv:1601.07512].

\bibitem{Patrignani:2016xqp}
  C.~Patrignani {\it et al.} [Particle Data Group],
  ``Review of Particle Physics,''
  Chin.\ Phys.\ C {\bf 40}, no. 10, 100001 (2016).

\bibitem{Vissani:1999tu}
  F.~Vissani,
  ``Signal of neutrinoless double beta decay, neutrino spectrum and oscillation scenarios,''
  JHEP {\bf 9906}, 022 (1999)
  [hep-ph/9906525].

\bibitem{Feruglio:2002af}
  F.~Feruglio, A.~Strumia and F.~Vissani,
  ``Neutrino oscillations and signals in beta and 0nu2beta experiments,''
  Nucl.\ Phys.\ B {\bf 637}, 345 (2002)
  Addendum: [Nucl.\ Phys.\ B {\bf 659}, 359 (2003)]
  [hep-ph/0201291].

\bibitem{Xing:2003jf}
  Z.~z.~Xing,
  ``Vanishing effective mass of the neutrinoless double beta decay?,''
  Phys.\ Rev.\ D {\bf 68}, 053002 (2003)
  [hep-ph/0305195].

\bibitem{KamLAND-Zen:2016pfg}
  A.~Gando {\it et al.} [KamLAND-Zen Collaboration],
  ``Search for Majorana Neutrinos near the Inverted Mass Hierarchy Region with KamLAND-Zen,''
  Phys.\ Rev.\ Lett.\  {\bf 117}, no. 8, 082503 (2016)
  Addendum: [Phys.\ Rev.\ Lett.\  {\bf 117}, no. 10, 109903 (2016)]
  [arXiv:1605.02889].

\bibitem{Licciardi:2017oqg}
  C.~Licciardi [nEXO Collaboration],
  ``The Sensitivity of the nEXO Experiment to Majorana Neutrinos,''
  J.\ Phys.\ Conf.\ Ser.\  {\bf 888}, no. 1, 012237 (2017).

\bibitem{Martin-Albo:2015rhw}
  J.~Mart\'{\i}n-Albo {\it et al.} [NEXT Collaboration],
  ``Sensitivity of NEXT-100 to Neutrinoless Double Beta Decay,''
  JHEP {\bf 1605}, 159 (2016)
  [arXiv:1511.09246].

\bibitem{Zhao:2016brs}
  J.~Zhao, L.~J.~Wen, Y.~F.~Wang and J.~Cao,
  ``Physics potential of searching for $0\nu\beta\beta$ decays in JUNO,''
  Chin.\ Phys.\ C {\bf 41}, no. 5, 053001 (2017)
  [arXiv:1610.07143].

\bibitem{Chen:2016qcd}
  X.~Chen {\it et al.},
  ``PandaX-III: Searching for neutrinoless double beta decay with high pressure $^{136}$Xe gas time projection chambers,''
  Sci.\ China Phys.\ Mech.\ Astron.\  {\bf 60}, no. 6, 061011 (2017)
  [arXiv:1610.08883].

\bibitem{Abgrall:2017syy}
  N.~Abgrall {\it et al.} [LEGEND Collaboration],
  ``The Large Enriched Germanium Experiment for Neutrinoless Double Beta Decay (LEGEND),''
  arXiv:1709.01980.

\bibitem{Xing:2015zha}
  Z.~z.~Xing, Z.~h.~Zhao and Y.~L.~Zhou,
  ``How to interpret a discovery or null result of the $0\nu 2\beta$ decay,''
  Eur.\ Phys.\ J.\ C {\bf 75}, no. 9, 423 (2015)
  [arXiv:1504.05820].

\bibitem{Xing:2016ymd}
  Z.~z.~Xing and Z.~h.~Zhao,
  ``The effective neutrino mass of neutrinoless double-beta decays: how possible to fall into a well,''
  Eur.\ Phys.\ J.\ C {\bf 77}, no. 3, 192 (2017)
  [arXiv:1612.08538].

\bibitem{Benato:2015via}
  G.~Benato,
  ``Effective Majorana Mass and Neutrinoless Double Beta Decay,''
  Eur.\ Phys.\ J.\ C {\bf 75}, no. 11, 563 (2015)
  [arXiv:1510.01089].

\bibitem{Ge:2016tfx}
  S.~F.~Ge and M.~Lindner,
  ``Extracting Majorana properties from strong bounds on neutrinoless double beta decay,''
  Phys.\ Rev.\ D {\bf 95}, no. 3, 033003 (2017)
  [arXiv:1608.01618].

\bibitem{Caldwell:2017mqu}
  A.~Caldwell, A.~Merle, O.~Schulz and M.~Totzauer,
  Phys.\ Rev.\ D {\bf 96}, no. 7, 073001 (2017)
  [arXiv:1705.01945].

\bibitem{Agostini:2017jim}
  M.~Agostini, G.~Benato and J.~Detwiler,
  ``Discovery probability of next-generation neutrinoless double-$\beta$ decay experiments,''
  Phys.\ Rev.\ D {\bf 96}, no. 5, 053001 (2017)
  [arXiv:1705.02996].

\bibitem{Ge:2017erv}
  S.~F.~Ge, W.~Rodejohann and K.~Zuber,
  Phys.\ Rev.\ D {\bf 96}, no. 5, 055019 (2017)
  [arXiv:1707.07904].

\bibitem{Gariazzo:2015rra}
  S.~Gariazzo, C.~Giunti, M.~Laveder, Y.~F.~Li and E.~M.~Zavanin,
  ``Light sterile neutrinos,''
  J.\ Phys.\ G {\bf 43}, 033001 (2016)
  [arXiv:1507.08204].

\bibitem{Adamson:2016jku}
  P.~Adamson {\it et al.} [Daya Bay and MINOS Collaborations],
  ``Limits on Active to Sterile Neutrino Oscillations from Disappearance Searches in the MINOS, Daya Bay, and Bugey-3 Experiments,''
  Phys.\ Rev.\ Lett.\  {\bf 117}, no. 15, 151801 (2016)
  Addendum: [Phys.\ Rev.\ Lett.\  {\bf 117}, no. 20, 209901 (2016)]
  [arXiv:1607.01177].

\bibitem{Gariazzo:2017fdh}
  S.~Gariazzo, C.~Giunti, M.~Laveder and Y.~F.~Li,
  ``Updated Global 3+1 Analysis of Short-BaseLine Neutrino Oscillations,''
  JHEP {\bf 1706}, 135 (2017)
  [arXiv:1703.00860].

\bibitem{Goswami:2005ng}
  S.~Goswami and W.~Rodejohann,
  ``Constraining mass spectra with sterile neutrinos from neutrinoless double beta decay, tritium beta decay and cosmology,''
  Phys.\ Rev.\ D {\bf 73}, 113003 (2006)
  [hep-ph/0512234].

\bibitem{Goswami:2007kv}
  S.~Goswami and W.~Rodejohann,
  ``MiniBooNE results and neutrino schemes with 2 sterile neutrinos: Possible mass orderings and observables related to neutrino masses,''
  JHEP {\bf 0710}, 073 (2007)
  [arXiv:0706.1462].

\bibitem{Barry:2011wb}
  J.~Barry, W.~Rodejohann and H.~Zhang,
  ``Light Sterile Neutrinos: Models and Phenomenology,''
  JHEP {\bf 1107}, 091 (2011)
  [arXiv:1105.3911].

\bibitem{Li:2011ss}
  Y.~F.~Li and S.~s.~Liu,
  ``Vanishing effective mass of the neutrinoless double beta decay including light sterile neutrinos,''
  Phys.\ Lett.\ B {\bf 706}, 406 (2012)
  [arXiv:1110.5795].

\bibitem{Girardi:2013zra}
  I.~Girardi, A.~Meroni and S.~T.~Petcov,
  ``Neutrinoless Double Beta Decay in the Presence of Light Sterile Neutrinos,''
  JHEP {\bf 1311}, 146 (2013)
  [arXiv:1308.5802].

\bibitem{Giunti:2015kza}
  C.~Giunti and E.~M.~Zavanin,
  ``Predictions for Neutrinoless Double-Beta Decay in the 3+1 Sterile Neutrino Scenario,''
  JHEP {\bf 1507}, 171 (2015)
  [arXiv:1505.00978].

\bibitem{Schechter:1981bd}
  J.~Schechter and J.~W.~F.~Valle,
  ``Neutrinoless Double beta Decay in SU(2) x U(1) Theories,''
  Phys.\ Rev.\ D {\bf 25}, 2951 (1982).

\bibitem{Duerr:2011zd}
  M.~Duerr, M.~Lindner and A.~Merle,
  ``On the Quantitative Impact of the Schechter-Valle Theorem,''
  JHEP {\bf 1106}, 091 (2011)
  [arXiv:1105.0901].

\bibitem{Liu:2016oph}
  J.~H.~Liu, J.~Zhang and S.~Zhou,
  ``Majorana Neutrino Masses from Neutrinoless Double-Beta Decays and Lepton-Number-Violating Meson Decays,''
  Phys.\ Lett.\ B {\bf 760}, 571 (2016)
  [arXiv:1606.04886].
\end{thebibliography}
\end{document}